\documentclass[onecolumn,showpacs,preprintnumbers]{revtex4}
\usepackage{mathrsfs}
\usepackage{graphicx}
\usepackage{dcolumn}
\usepackage{bm}
\usepackage{epsfig}
\usepackage{amsmath,amssymb}

\begin{document}
\title{Analysis of the $X(3842)$ as a D-wave charmonium meson}
\author{Guo-Liang Yu$^{1}$}
\email{yuguoliang2011@163.com}
\author{Zhi-Gang Wang$^{1}$}
\email{zgwang@aliyun.com}

\affiliation{Department of Mathematics and Physics, North China
Electric power university, Baoding 071003, People's Republic of
China}
\date{\today }

\begin{abstract}
In this article, we assign the newly reported state $X(3842)$ to be a D-wave $\overline{c}c$ meson, and study its mass and decay
constant with the QCD sum rules by considering the contributions of the vacuum condensates up to dimension-6 in the operator
product expansion. The predicted mass $M_{X(3842)}=(3.844^{+0.0675}_{-0.0823}\pm0.020)GeV$
is in agreement well with the experimental data $M_{X(3842)}=(3842.71\pm0.16\pm0.12)MeV$ from the LHCb collaboration. This result
supports assigning $X(3842)$ to be a $1^{3}D_{3}$ charmonium meson. As the $1^{3}D_{3}$ $\overline{c}c$ meson, its predicted strong decay width with
the $^{3}P_{0}$ decay model is compatible with the experimental data.
\end{abstract}


\maketitle

\begin{large}
\textbf{1 Introduction}
\end{large}

Since the observation of the $J/\psi$ resonance in 1974\cite{Aubert,Augustin}, theoretical and experimental physicist have mapped out the spectrum of hidden charm mesons
with high precision. The experimentally clear spectrum of relatively narrow states below the open-charm $DD$ threshold
of $3.73GeV$ have been identified with the $1S$, $1P$, $2S$ charmonium states. In 2003, the Belle collaboration reported a new
charmonium-like state $X(3872)$\cite{Choi}, which was recently assigned to be the $\chi_{c1}$ meson after some controversy\cite{Tanabashi}. Subsequently, some other charmonium-like states
such as $X(3860)$\cite{Chilikin}, $X(3915)$\cite{Lees}, $X(3930)$\cite{Uehara,Aubert1}, $X(3940)$\cite{Abe,Pakhlov} were reported by Belle, BARBAR collaborations. These states do not fit into the conventional
hidden-charm spectrum and are believed to be exotic in nature. These states were explained to be different structures such as
a charmonium state\cite{Tanabashi,Colangelo,Fulvia,Braguta,LiuX,Lees1}, a molecule state\cite{mole1,mole2,mole3,mole4,mole5,mole6,mole7,mole8,mole9,mole10}, a tetraquark state\cite{WZG1,tetro1,tetro2,tetro3,tetro4,tetro5} or a mixture of charmonium and molecular $DD^{*}$ component\cite{Albuquerque,Fernandez}.
Stimulated by these new exotic states, there has been a resurgence of interest in charmonium spectroscopy.

The expected charmonium states $\eta_{c2}(1^{1}D_{2})$ and $\psi_{3}(1^{3}D_{3})$\cite{Eichten1,Eichten2}, which is close to $D\overline{D}$ threshold, still remain undiscovered in experiment.
Though the $\psi_{3}(1^{3}D_{3})$ state lies above the open charm
threshold, the decay channel to the $D\overline{D}$ is suppressed due to the F-wave centrifugal barrier factor.
Consequently, the $\psi_{3}(1^{3}D_{3})$ state is expected to be narrow with
a natural width of $1\sim2MeV$\cite{Barnes1,Barnes2}. Predictions for the mass of this state lie in the range
$3815\sim3863 MeV/c^{2}$\cite{Eichten2,pred1,pred2,pred3,pred4,pred5,pred6,pred7}.

Very recently, LHCb collaboration studied the near-threshold $D\overline{D}$ mass spectra using the LHCb dataset collected between 2011 and 2018
and observed a new narrow charmonium state in the decay modes $X(3842)\rightarrow D^{0}\overline{D^{0}}$ and
$X(3842)\rightarrow D^{+}D^{-}$ with very high statistical significance\cite{Aaij}. The mass and the natural width of this state are measured to be
\begin{eqnarray}
\notag
M_{X(3842)}=(3842.71\pm0.16\pm0.12)MeV
\end{eqnarray}
\begin{eqnarray}
\notag
\Gamma_{X(3842)}=(2.79\pm0.51\pm0.35)MeV
\end{eqnarray}
The narrow natural width and measured value of the mass suggests the interpretation of the $X(3842)$ state as
the $\psi_{3}(1^{3}D_{3})$ charmonium state with $J^{PC} = 3^{--}$.

In order to make a further conformation about the nature of the $X(3842)$, we calculate the mass of this charmonium state
based on QCD sum rules. QCD sum rules proved to be a most powerful theoretical tool in studying the ground state hadrons and
it has been widely used to analyze the masses, decay constants, form factors and strong coupling constants, etc\cite{Shifman,Reinders}.
There have been many reports about the spin-parity $J^{PC}=0^{\pm}$, $1^{\pm}$ mesons with the QCD sum rules\cite{WZG2,Narison}, while the
works on the $J^{PC}=2^{+}$, $3^{-}$ are few\cite{Sundu,WZG3}. The $^{3}P_{0}$ decay model is an effective
and simple method, which can give good description about the strong decay behaviors of many
hadrons\cite{Blunder,ZhouHQ,LiDM,ZhanbB,GuoLY}. In this article, we assgin the $X(3842)$ to be a D-wave $\overline{c}c$ meson, study
its mass and decay constant with the full QCD sum rules in detail by considering the contributions of the vacuum condensates
up to dimension-6 in the operator expansion and calculate the strong decay width of the $X(3842)$ with $^{3}P_{0}$ decay model.

The layout of this paper is as follows: we derive the QCD sum rules for the mass and decay constant of the $X(3842)$ and present
the numerical results in Sec.2; in Sec.3, we analyze the strong decay width with the $^{3}P_{0}$ decay model; and Sec.4 is reserved
for our conclusions.

\begin{large}
\textbf{2 QCD sum rules for $X(3842)$ as a $1^{3}D_{3}$  charmonium state}
\end{large}

To study the mass and decay constant of X(3842), we first write down the following
two-point correlation function,
\begin{eqnarray}
\Pi_{\mu\nu\rho\alpha\beta\sigma}(p)=i\int d^{4}xe^{ip.(x-y)}\Big\langle0|\mathbb{T}\Big\{J_{\mu\nu\rho}(x)J_{\alpha\beta\sigma}^{\dag}(y)\Big\}
|0\Big\rangle,
\end{eqnarray}
where $\mathbb{T}$ is the time ordered product and
$J$ is
the interpolating current of $X(3842)$. The interpolating
current is a composite operator with the same quantum numbers as the
studied hadron. In this work, the current can be written as,
\begin{eqnarray}
&& J_{\mu\nu\rho}(x)=\overline{c}(x)\Big(\gamma_{\mu}\overleftrightarrow{D}_{\nu}\overleftrightarrow{D}_{\rho}+
\gamma_{\nu}\overleftrightarrow{D}_{\rho}\overleftrightarrow{D}_{\mu}+\gamma_{\rho}\overleftrightarrow{D}_{\mu}\overleftrightarrow{D}_{\nu} \Big)c(x)
\end{eqnarray}
with $\overleftrightarrow{D}_{\mu}=(\overrightarrow{\partial}_{\mu}-ig_{s}G_{\mu})-(\overleftarrow{\partial}_{\mu}+ig_{s}G_{\mu})$, where
$D_{\mu}$, $\partial_{\mu}$ are the covariant derivative, partial derivative and $G_{\mu}$ is the gluon field. This current can be decomposed
into two parts,
\begin{eqnarray}
J_{\mu\nu\rho}(x)=\eta_{\mu\nu\rho}(x)+J^{V}_{\mu\nu\rho}(x)
\end{eqnarray}
where
\begin{eqnarray}
\eta_{\mu\nu\rho}(x)=&&\overline{c}(x)\Big(\gamma_{\mu}\overleftrightarrow{\partial}_{\nu}\overleftrightarrow{\partial}_{\rho}+
\gamma_{\nu}\overleftrightarrow{\partial}_{\rho}\overleftrightarrow{\partial}_{\mu}+\gamma_{\rho}\overleftrightarrow{\partial}_{\mu}\overleftrightarrow{\partial}_{\nu} \Big)c(x) \\
\notag
J^{V}_{\mu\nu\rho}(x)=&&-2i\overline{c}(x)\Big[\gamma_{\mu}\Big(g_{s}G_{\nu}\overleftrightarrow{\partial}_{\rho}+\overleftrightarrow{\partial}_{\nu}g_{s}G_{\rho}-2ig_{s}^{2}G_{\nu}G_{\rho}\Big) \\
&& \notag
+\gamma_{\nu}\Big(g_{s}G_{\rho}\overleftrightarrow{\partial}_{\mu}+\overleftrightarrow{\partial}_{\rho}g_{s}G_{\mu}-2ig_{s}^{2}G_{\rho}G_{\mu}\Big) \\
&&
+\gamma_{\rho}\Big(g_{s}G_{\mu}\overleftrightarrow{\partial}_{\nu}+\overleftrightarrow{\partial}_{\mu}g_{s}G_{\nu}-2ig_{s}^{2}G_{\mu}G_{\nu}\Big) \Big]c(x)
\end{eqnarray}
and $\overleftrightarrow{\partial}_{\mu}=\overrightarrow{\partial}_{\mu}-\overleftarrow{\partial}_{\mu}$. The current $J_{\mu\nu\rho}(x)$ of Eq.(2) is constructed with covariant derivative
$D_{\mu}$ which is gauge invariant, but blurs the physical interpretation of the $\overleftrightarrow{\partial}_{\mu}$ being the angular. The current $\eta_{\mu\nu\rho}(x)$ of Eq.(4) with the partial
derivative $\partial_{\mu}$ destroy the invariance of gauge transformation, but manifests the physical interpretation of $\overleftrightarrow{\partial}_{\mu}$ being the angular momentum.
In this work, we will present the results which are obtained from these two currents $J_{\mu\nu\rho}(x)$ and $\eta_{\mu\nu\rho}(x)$ separately.

\begin{large}
\textbf{2.1 The phenomenological side}
\end{large}

In order to obtain the phenomenological representations, we insert a
complete set of intermediate hadronic states with the same quantum
numbers as the current operators
$J_{\mu\nu\rho}(x)$ into
the correlation $\Pi_{\mu\nu\rho\alpha\beta\sigma}(p)$\cite{Shifman,Reinders}. It should be noticed that the current $J_{\mu\nu\rho}(0)$ has negative parity,
and couples potentially to the $J^{p}=3^{-}$, $2^{+}$, $1^{-}$, $0^{+}$ $\overline{c}c$ mesons,
\begin{eqnarray}
\notag
&&\langle 0| J_{\mu\nu\rho}(0)|X(3842)(p) \rangle=f_{X(3842)}\varepsilon_{\mu\nu\rho}(p,\lambda),\\
\notag
&&\langle 0| J_{\mu\nu\rho}(0)|\chi_{c2}(p) \rangle=f_{\chi_{c2}}[p_{\mu}\varepsilon_{\nu\rho}(p,\lambda)+p_{\nu}\varepsilon_{\rho\mu}(p,\lambda)+p_{\rho}\varepsilon_{\mu\nu}(p,\lambda)],\\
\notag
&&\langle 0| J_{\mu\nu\rho}(0)|J/\psi(p) \rangle=f_{J/\psi}[p_{\mu}p_{\nu}\varepsilon_{\rho}(p,\lambda)+p_{\nu}p_{\rho}\varepsilon_{\mu}(p,\lambda)+p_{\rho}p_{\mu}\varepsilon_{\nu}(p,\lambda)],\\
\notag
&&\langle 0| J_{\mu\nu\rho}(0)|\chi_{c0}(p) \rangle=f_{\chi_{c0}}p_{\mu}p_{\nu}p_{\rho}(p,\lambda),
\end{eqnarray}
where $f$ are the decay constants, and $\varepsilon$ are the polarization vectors of the $\overline{c}c$ mesons with the following properties\cite{ZhuJJ},
\begin{eqnarray}
\notag
P_{\mu\nu\rho\alpha\beta\sigma}&&= \mathop{\sum}\limits_\lambda\varepsilon_{\mu\nu\rho}^{*}(\lambda,p)\varepsilon_{\alpha\beta\sigma}(\lambda,p) \\
\notag
&&=\frac{1}{6}(\widetilde{g}_{\mu\alpha}\widetilde{g}_{\nu\beta}\widetilde{g}_{\rho\sigma}+\widetilde{g}_{\mu\alpha}\widetilde{g}_{\nu\sigma}\widetilde{g}_{\rho\beta}
+\widetilde{g}_{\mu\beta}\widetilde{g}_{\nu\alpha}\widetilde{g}_{\rho\sigma}+\widetilde{g}_{\mu\beta}\widetilde{g}_{\nu\sigma}\widetilde{g}_{\rho\alpha}+
\widetilde{g}_{\mu\sigma}\widetilde{g}_{\nu\alpha}\widetilde{g}_{\rho\beta}+\widetilde{g}_{\mu\sigma}\widetilde{g}_{\nu\beta}\widetilde{g}_{\rho\alpha})\\
\notag
&&
-\frac{1}{15}(\widetilde{g}_{\mu\alpha}\widetilde{g}_{\nu\rho}\widetilde{g}_{\beta\sigma}+\widetilde{g}_{\mu\beta}\widetilde{g}_{\nu\rho}\widetilde{g}_{\alpha\sigma}+
\widetilde{g}_{\mu\sigma}\widetilde{g}_{\nu\rho}\widetilde{g}_{\alpha\beta}+\widetilde{g}_{\nu\alpha}\widetilde{g}_{\mu\rho}\widetilde{g}_{\beta\sigma}
+\widetilde{g}_{\nu\beta}\widetilde{g}_{\mu\rho}\widetilde{g}_{\alpha\sigma}+\widetilde{g}_{\nu\sigma}\widetilde{g}_{\mu\rho}\widetilde{g}_{\alpha\beta}\\
&&+\widetilde{g}_{\rho\alpha}\widetilde{g}_{\mu\nu}\widetilde{g}_{\beta\sigma}+\widetilde{g}_{\rho\beta}\widetilde{g}_{\mu\nu}\widetilde{g}_{\alpha\sigma}
+\widetilde{g}_{\rho\sigma}\widetilde{g}_{\mu\nu}\widetilde{g}_{\alpha\beta})
\end{eqnarray}
\begin{eqnarray}
P_{\mu\nu\alpha\beta}= \mathop{\sum}\limits_\lambda\varepsilon_{\mu\nu}^{*}(\lambda,p)\varepsilon_{\alpha\beta}(\lambda,p)
=\frac{\widetilde{g}_{\mu\alpha}\widetilde{g}_{\nu\beta}+\widetilde{g}_{\mu\beta}\widetilde{g}_{\nu\alpha}}{2}-\frac{\widetilde{g}_{\mu\nu}\widetilde{g}_{\alpha\beta}}{3}
\end{eqnarray}
\begin{eqnarray}
\widetilde{g}_{\mu\nu}= \mathop{\sum}\limits_\lambda\varepsilon_{\mu}^{*}(\lambda,p)\varepsilon_{\nu}(\lambda,p)
=-g_{\mu\nu}+\frac{p_{\mu}p_{\nu}}{p^{2}}
\end{eqnarray}
With these above equations, the correlation function can be decomposed into the following structures,
\begin{eqnarray}
\notag
\Pi_{\mu\nu\rho\alpha\beta\sigma}(p)&&=\Pi(p^{2})P_{\mu\nu\rho\alpha\beta\sigma}+\Pi_{2}(p^{2})\Big[P_{\nu\rho\beta\sigma}p_{\mu}p_{\alpha}+P_{\nu\rho\alpha\sigma}p_{\mu}p_{\beta}
+P_{\nu\rho\alpha\beta}p_{\mu}p_{\sigma}+P_{\mu\rho\beta\sigma}p_{\nu}p_{\alpha}\\
\notag
&&+P_{\mu\rho\alpha\sigma}p_{\nu}p_{\beta}+P_{\mu\rho\alpha\beta}p_{\nu}p_{\sigma}+P_{\mu\nu\beta\sigma}p_{\rho}p_{\alpha}+P_{\mu\nu\alpha\sigma}p_{\rho}p_{\beta}+P_{\mu\nu\alpha\beta}p_{\rho}p_{\sigma}\Big]\\
\notag
&&+\Pi_{1}(p^{2})\Big[\widetilde{g}_{\mu\alpha}p_{\nu}p_{\rho}p_{\beta}p_{\sigma}+\widetilde{g}_{\mu\beta}p_{\nu}p_{\rho}p_{\alpha}p_{\sigma}+\widetilde{g}_{\mu\sigma}p_{\nu}p_{\rho}p_{\alpha}p_{\beta}+
\widetilde{g}_{\nu\alpha}p_{\mu}p_{\rho}p_{\beta}p_{\sigma}\\
\notag &&+\widetilde{g}_{\nu\beta}p_{\mu}p_{\rho}p_{\alpha}p_{\sigma}+\widetilde{g}_{\nu\sigma}p_{\mu}p_{\rho}p_{\alpha}p_{\beta}+\widetilde{g}_{\rho\alpha}p_{\mu}p_{\nu}p_{\beta}p_{\sigma}
+\widetilde{g}_{\rho\beta}p_{\mu}p_{\nu}p_{\alpha}p_{\sigma}+\widetilde{g}_{\rho\sigma}p_{\mu}p_{\nu}p_{\alpha}p_{\beta}\Big]\\
&&+\Pi_{0}(p^{2})p_{\mu}p_{\nu}p_{\rho}p_{\alpha}p_{\beta}p_{\sigma}
\end{eqnarray}
Here, the component $\Pi(p^{2})$ denotes the contribution of the $1^{3}D_{3}$ charmonium state and it can be extracted out from the correlation function by the following projection method,
\begin{eqnarray}
\Pi(p^{2})&&=\frac{1}{7}P^{\mu\nu\rho\alpha\beta\sigma}\Pi_{\mu\nu\rho\alpha\beta\sigma}(p)
\end{eqnarray}
according to the properties,
\begin{eqnarray}
p^{\mu}P_{\mu\nu\rho\alpha\beta\sigma}=p^{\nu}P_{\mu\nu\rho\alpha\beta\sigma}=p^{\rho}P_{\mu\nu\rho\alpha\beta\sigma}
=p^{\alpha}P_{\mu\nu\rho\alpha\beta\sigma}=p^{\beta}P_{\mu\nu\rho\alpha\beta\sigma}=p^{\sigma}P_{\mu\nu\rho\alpha\beta\sigma}=0
\end{eqnarray}
After the ground-state
contribution is isolated, we get the following function,
\begin{eqnarray}
\notag\
\Pi_{\mu\nu\rho\alpha\beta\sigma}(p)&&=\frac{f_{X(3842)}^2}{M_{X(3842)}^2-p^2}P_{\mu\nu\rho\alpha\beta\sigma}+\cdots \\
&& =\Pi(p^2)P_{\mu\nu\rho\alpha\beta\sigma}+\cdots
\end{eqnarray}
where $h.r.$ stands for the contributions of the higher resonances
and continuum states.

\begin{large}
\textbf{2.2 The OPE side}
\end{large}

Considering all possible contractions of the quark fields with
Wick's theorem, the correlation function Eq.$(1)$ is written as
\begin{eqnarray}
\Pi(p^{2})=&&-\frac{i}{7}P^{\mu\nu\rho\alpha\beta\sigma}\int d^{4}xe^{ip.(x-y)}\times\Big\{\Gamma^{ik}_{\mu\nu\rho}(x)S^{c}_{kl}(x-y)\Gamma_{\alpha\beta\sigma}^{lj}(y)S_{ji}^{c}(y-x) \Big\}|_{y=0}
\end{eqnarray}
where $\Gamma_{\mu\nu\rho}(x)$ and $\Gamma_{\alpha\beta\sigma}(y)$ are the vertexes, $S$ are the quark propagators which are replaced with the
following 'full' propagators,
\begin{eqnarray}
\notag
 S_{c}^{mn}(x)=&&\frac{i}{(2\pi)^{4}}\int
 d^{4}ke^{-ik.x}\Big\{\frac{\delta_{mn}}{k\!\!\!/-m_{c}}-\frac{g_{s}G_{\alpha\beta}^{c}t^{c}_{mn}}{4}\frac{\sigma^{\alpha\beta}(k\!\!\!/+m_{c})+(k\!\!\!/+m_{c})\sigma^{\alpha\beta}}{(k^{2}-m_{c}^{2})^{2}}\\
 &&-\frac{g_{s}^{2}(t^{a}t^{b})_{mn}G_{\alpha\beta}^{a}G_{\mu\nu}^{b}(f^{\alpha\beta\mu\nu}+f^{\alpha\mu\beta\nu}+f^{\alpha\mu\nu\beta})}{4(k^{2}-m_{c}^{2})^{5}}\\
 \notag &&
 + \frac{i\langle g_{s}^{3}GGG\rangle}{48}\frac{(k\!\!\!/+m_{c})\big[k\!\!\!/(k^{2}-3m_{c}^{2})+2m_{c}(2k^{2}-m_{c}^{2})\big](k\!\!\!/+m_{c})}{(k^{2}-m_{c}^{2})^{6}}\cdots\Big\}
\end{eqnarray}
where
\begin{eqnarray}
f^{\alpha\beta\mu\nu}=(k\!\!\!/+m_{c})\gamma^{\alpha}(k\!\!\!/+m_{c})\gamma^{\beta}(k\!\!\!/+m_{c})\gamma^{\mu}(k\!\!\!/+m_{c})\gamma^{\nu}(k\!\!\!/+m_{c})
\end{eqnarray}
$t^{n}=\frac{\lambda^{n}}{2}$, the $\lambda^{n}$ is the Gell-Mann matrix, $m$, $n$ are color indexes. In the fixed point gauge, $G_{\mu}(x)=\frac{1}{2}x^{\theta}G_{\theta\mu}(0)+\cdots$ and $G_{\alpha}(y)=\frac{1}{2}y^{\theta}G_{\theta\alpha}(0)+\cdots$=0.
Thus, for the vertex $\Gamma_{\alpha\beta\sigma}(y)$, we can get $G_{\alpha}(y)=G_{\beta}(y)=G_{\sigma}(y)=0$, there are no gluon lines associated with the vertex at the point $y=0$.
Then we complete the integrals both in the coordinate and momentum spaces and obtain the QCD spectral density through dispersion relation,
\begin{eqnarray}
\Pi(p^{2})=\frac{1}{\pi}\int^{\infty}_{4m_{c}^{2}}\frac{Im\Pi(s)}{s-p^{2}}ds=\int^{\infty}_{4m_{c}^{2}}\frac{\rho_{QCD}(s)}{s-p^{2}}ds
\end{eqnarray}

where the contributions of the perturbative terms, $\langle\frac{\alpha_{s}GG}{\pi}\rangle$, $\langle g_{s}^{3}GGG\rangle$ are written as,
\begin{eqnarray}
\rho_{0}^{\eta}(s)=\frac{9}{35\pi^{2}}\times\frac{(s-4m_{c}^{2})^{\frac{5}{2}}(s+3m_{c}^{2})}{\sqrt{s}}exp\Big(-\frac{s}{T^2}\Big)
\end{eqnarray}
\begin{eqnarray}
\rho_{GG}^{\eta}(s)=\langle\frac{\alpha_{s}GG}{\pi}\rangle \times\Big[\frac{6(32m_{c}^{6}-22sm_{c}^{4}+3s^{2}m_{c}^2)}{4s\sqrt{s(s-4m_{c}^{2})}}+\frac{\sqrt{s(s-4m_{c}^{2})}(22m_{c}^{4}+19sm_{c}^{2}-8s^2)}{10s^2} \Big]
\end{eqnarray}
\begin{eqnarray}
\notag
\rho_{GGG}^{\eta}(s)=&&\langle g_{s}^{3}GGG\rangle \times\Big[\frac{6(-192m_{c}^{8}+496sm^{6}_{c}-196s^{2}m^{4}_{c}+18s^{3}m^{2}_{c}+s^{4})}{64\pi^{2}s^{2}(4m^{2}_{c}-s)\sqrt{s(s-4m_{c}^{2})}} \\
&& +\frac{2(-576m_{c}^{8}+560sm^{6}_{c}-44s^{2}m^{4}_{c}-42s^{3}m^{2}_{c}+7s^{4})}{64\pi^{2}s^{2}(4m^{2}_{c}-s)\sqrt{s(s-4m_{c}^{2})}} \Big]
\end{eqnarray}
\begin{eqnarray}
\notag
\rho^{V}_{GG}(s)=&&\langle\frac{\alpha_{s}GG}{\pi}\rangle \times\Big[-\frac{(s-4m_{c}^{2})^{\frac{3}{2}}}{8\pi^{2}\sqrt{s}}-\frac{\sqrt{s(s-4m_{c}^{2})}(s+2m_{c}^{2})}{8\pi^{2}s^{2}}-\frac{\big[s(s-4m_{c}^{2})\big]^{\frac{3}{2}}}{8\pi^{2}s^{2}} \\
&&-\frac{\sqrt{s(s-4m_{c}^{2})}(s+2m_{c}^{2})}{8\pi^{2}s^{2}}+\frac{3\sqrt{s(s-4m_{c}^{2})}(s^{2}+2m_{c}^{2}s-4m_{c}^{4})}{40\pi^{2}s^{2}}\Big]
\end{eqnarray}
\begin{eqnarray}
\notag &&
\rho^{V}_{GGG}(s)=\langle g_{s}^{3}GGG\rangle \times\Big[\frac{(s+2m_{c}^{2})\sqrt{s(s-4m_{c}^{2})}}{8\pi^{2}s^{2}}+\frac{3m_{c}^{4}}{4\pi^{2}s\sqrt{s(s-4m_{c}^{2})}}
+\frac{-7m_{c}^{4}+6sm_{c}^{2}-s^{2}}{4\pi^{2}s\sqrt{s(s-4m_{c}^{2})}}\\
 &&
+\frac{256m_{c}^{6}-192m_{c}^{4}s+38m_{c}^{2}s^{2}-s^{3}}{32\pi^{2}s(4m_{c}^{2}-s)\sqrt{s(s-4m_{c}^{2})}}+\frac{(m_{c}^{2}s-3m_{c}^{4})}{4\pi^{2}s\sqrt{s(s-4m_{c}^{2})}}
+\frac{32m_{c}^{6}-24m_{c}^{4}s+2m_{c}^{2}s^{2}+s^{3}}{32\pi^{2}s(4m_{c}^{2}-s)\sqrt{s(s-4m_{c}^{2})}}\Big]
\end{eqnarray}
Here, the densities $\rho^{\eta}$, $\rho^{V}$ denote contributions coming from currents of $\eta_{\mu\nu\rho}$ and $J_{\mu\nu\rho}^{V}$ respectively.
To obtain the physical parameters of the hadron, we need to take quark-hadron duality below the threshold $s_{0}$ and perform the Borel transform with
respect to the variable $P^{2}=-p^{2}$,
\begin{eqnarray}
f_{X(3842)}^{2}exp\Big[-\frac{M_{X(3842)}^2}{T^2}\Big]=\int_{4m_{c}^{2}}^{s_{0}}\rho_{QCD}(s)exp\Big[-\frac{s}{T^2}\Big]ds
\end{eqnarray}
We differentiate Eq.(22) with respect to $\frac{1}{T^{2}}$, then eliminate the decay constant $f_{X(3842)}$, and obtain
\begin{eqnarray}
M_{X(3842)}^{2}=-\frac{\frac{d}{d(1/T^{2})}B_{T^2}\Pi(p^{2})}{B_{T^{2}}\Pi(p^{2})}
\end{eqnarray}
After the mass $M_{X(3842)}$ is obtained, it is treated as a input parameter to obtain the decay constant from QCD sum from Eq.(22).

\begin{large}
\textbf{2.3 The numerical results}
\end{large}

To obtain the physical parameters according to QCD sum rule, we need to determine the values of a few input parameters, such as
$\langle\frac{\alpha_{s}GG}{\pi}\rangle$, $\langle g_{s}^{3}GGG\rangle$, the mass of $m_{c}$ and threshold parameter $s_{0}$.
The values of $\langle\frac{\alpha_{s}GG}{\pi}\rangle$ and $\langle g_{s}^{3}GGG\rangle$ have been updated from time to time, and change
greatly. The recently updated values $\langle\frac{\alpha_{s}GG}{\pi}\rangle=(0.022\pm0.004)GeV^{4}$\cite{Narison22} and the three-gluon condensate
$\langle g_{s}^{3}GGG\rangle=(0.616\pm0.385)GeV^{6}$\cite{Narison22}, which have changed a lot comparing with the previous standard values
$\langle\frac{\alpha_{s}GG}{\pi}\rangle=0.012GeV^{4}$ and $\langle g_{s}^{3}GGG\rangle=0.045GeV^{6}$\cite{Shifman,Reinders,Colangelo22}.
In this wok, we take the updated values of the gluon condensate and three gluon condensate as the input. As for the quark mass is concerned, we can take the
$\overline{MS}$ mass $m_{c}(m_{c})=(1.275\pm0.025)GeV$ from the Particle Data Group\cite{Tanabashi}. According to the renormalization group equation,
the $\overline{MS}$ mass has the energy-dependence and is called the "running" mass\cite{Tanabashi},
\begin{eqnarray}
\notag
m_{c}(\mu)&&=m_{c}(m_{c})\Big[\frac{\alpha_{s}(\mu)}{\alpha_{s}(m_{c})}\Big]^{\frac{12}{25}} \\
\notag
\alpha_{s}(\mu)&&=\frac{1}{b_{0}t}\Big[ 1-\frac{b_{1}}{b_{0}^{2}}\frac{logt}{t}+\frac{b_{1}^{2}(log^{2}t-logt-1)+b_{0}b_{2}}{b_{0}^{4}t^{2}}\Big]
\end{eqnarray}
In addition, we can also take the pole mass which relates with the $\overline{MS}$ mass through the following relation\cite{Tanabashi},
\begin{eqnarray}
\notag
m_{c}=m_{c}(m_{c})\Big[1+\frac{4\alpha_{s}(m_{c})}{3\pi}+\cdots\Big]
\end{eqnarray}
In our previous work, we have discussed this problem in detail, which showed that $\overline{MS}$ and pole mass lead to little difference of the results\cite{WZG33}.
In this work, we take the pole mass $m_{c}=(1.275\pm0.025)GeV$.

Commonly, the energy gap between the ground state and the first radial excited state is about $0.5MeV$ for a conventional meson. Considering the measured
mass and width of $X(3842)$, which are $m_{X(3842)}=(3842.71\pm0.16\pm0.12)MeV$ and $\Gamma_{X(3842)}=(2.79\pm0.51\pm0.35)MeV$, we take the threshold parameter
$\sqrt{s_{0}}=3.84+(0.4\sim0.6)GeV$ to avoid the contaminations of the high resonances and continuum states.
It can also be seen from Eqs.$(22)$ and $(23)$ that the mass or the decay constant is the function of the
Borel parameters $T^{2}$. We commonly search for the optimal values of the Borel parameters basing on two
considerations which are pole dominance and convergence of the operator product expansion. That is to say, the pole contribution should be as large as
possible(larger than $40\%$) comparing with the contributions of the high resonances and continuum states. Meanwhile, we should also find a plateau, which will ensure
operator product expansion convergence and the stability of our results. The plateau is often called "Borel window".

\begin{figure}[h]
\begin{minipage}[t]{0.45\linewidth}
\centering
\includegraphics[height=5cm,width=7cm]{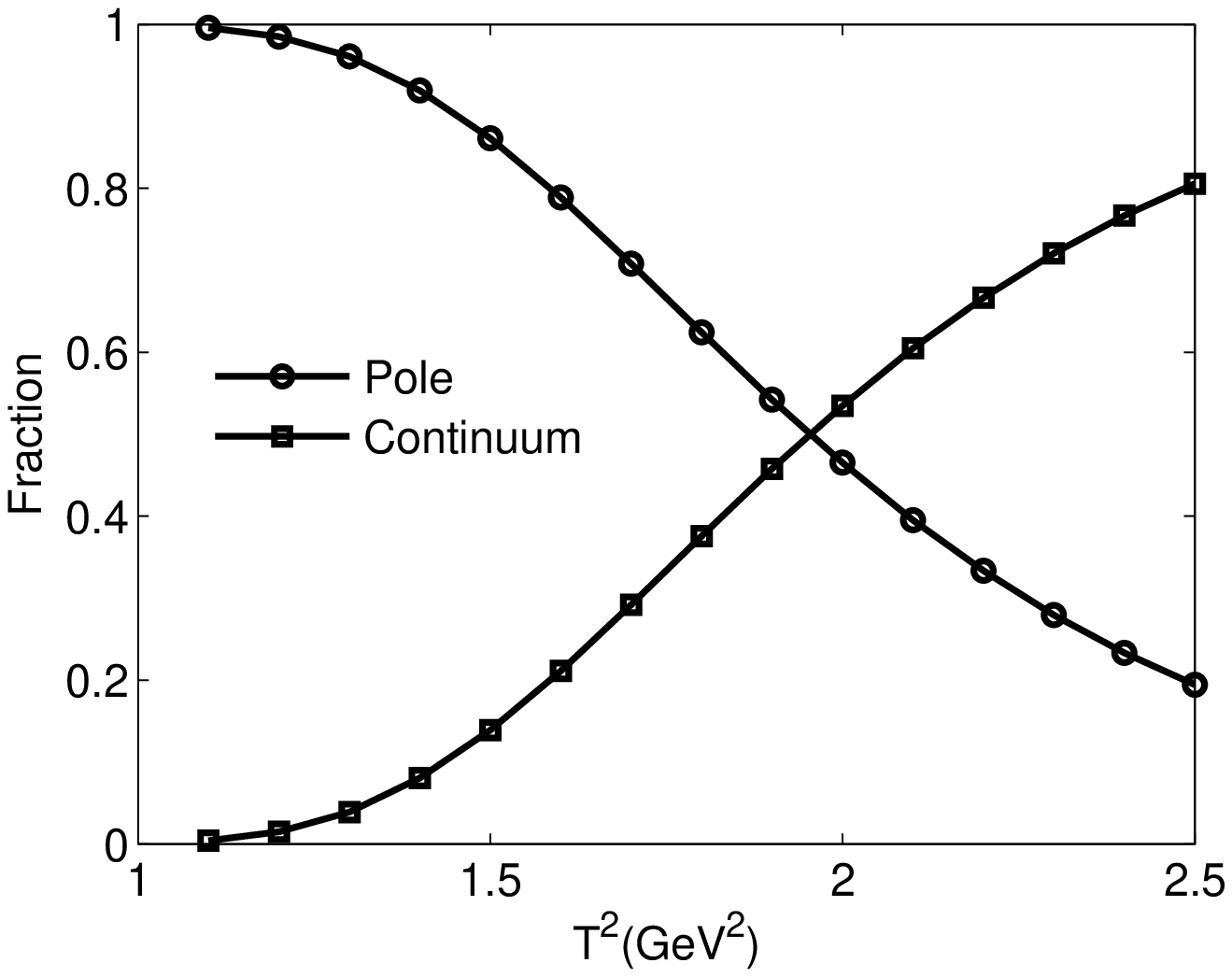}
\caption{The contributions come from pole and continuum state with variations of the Borel
parameter $T^2$.\label{your
label}}
\end{minipage}
\hfill
\begin{minipage}[t]{0.45\linewidth}
\centering
\includegraphics[height=5cm,width=7cm]{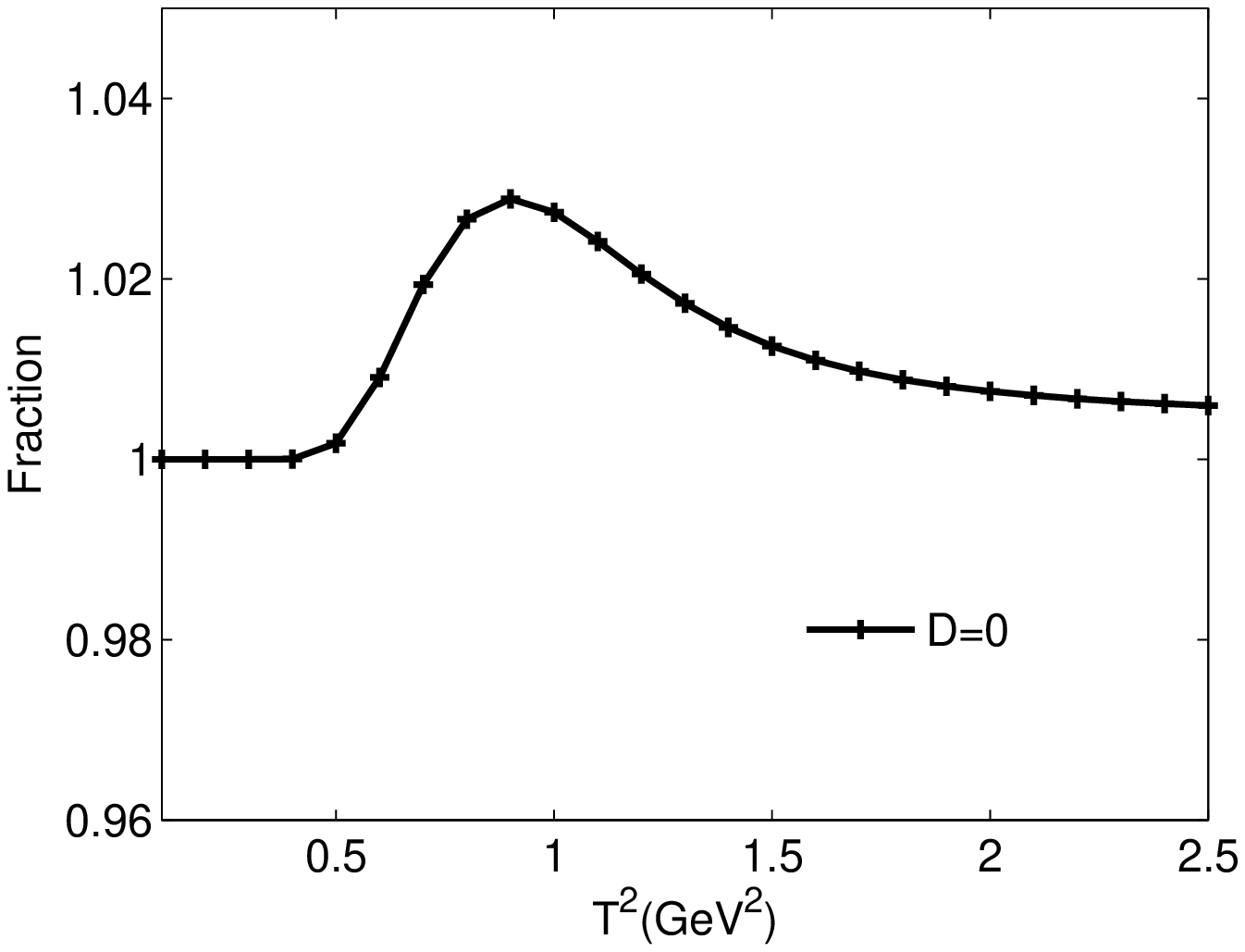}
\caption{The contribution come from purturbative term with variations of the Borel
parameter $T^2$.\label{your
label}}
\end{minipage}
\end{figure}
\begin{figure}[h]
\begin{minipage}[t]{0.45\linewidth}
\centering
\includegraphics[height=5cm,width=7cm]{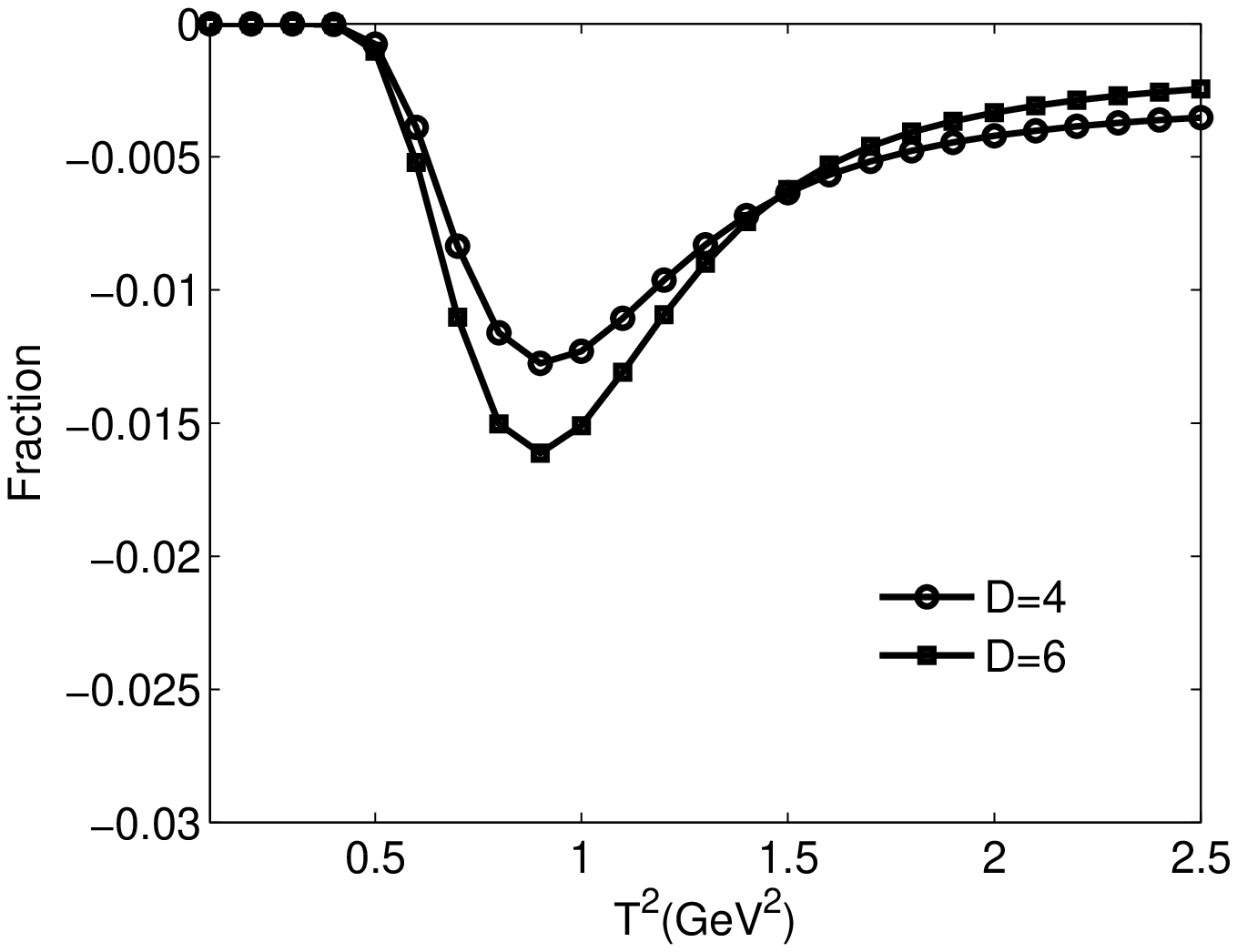}
\caption{The contributions come from $\langle\frac{\alpha_{s}}{\pi}GG\rangle$ and $\langle g_{s}^{3}GGG\rangle$
with variations of the Borel parameter $T^2$.\label{your label}}
\end{minipage}
\hfill
\begin{minipage}[t]{0.45\linewidth}
\centering
\includegraphics[height=5cm,width=7cm]{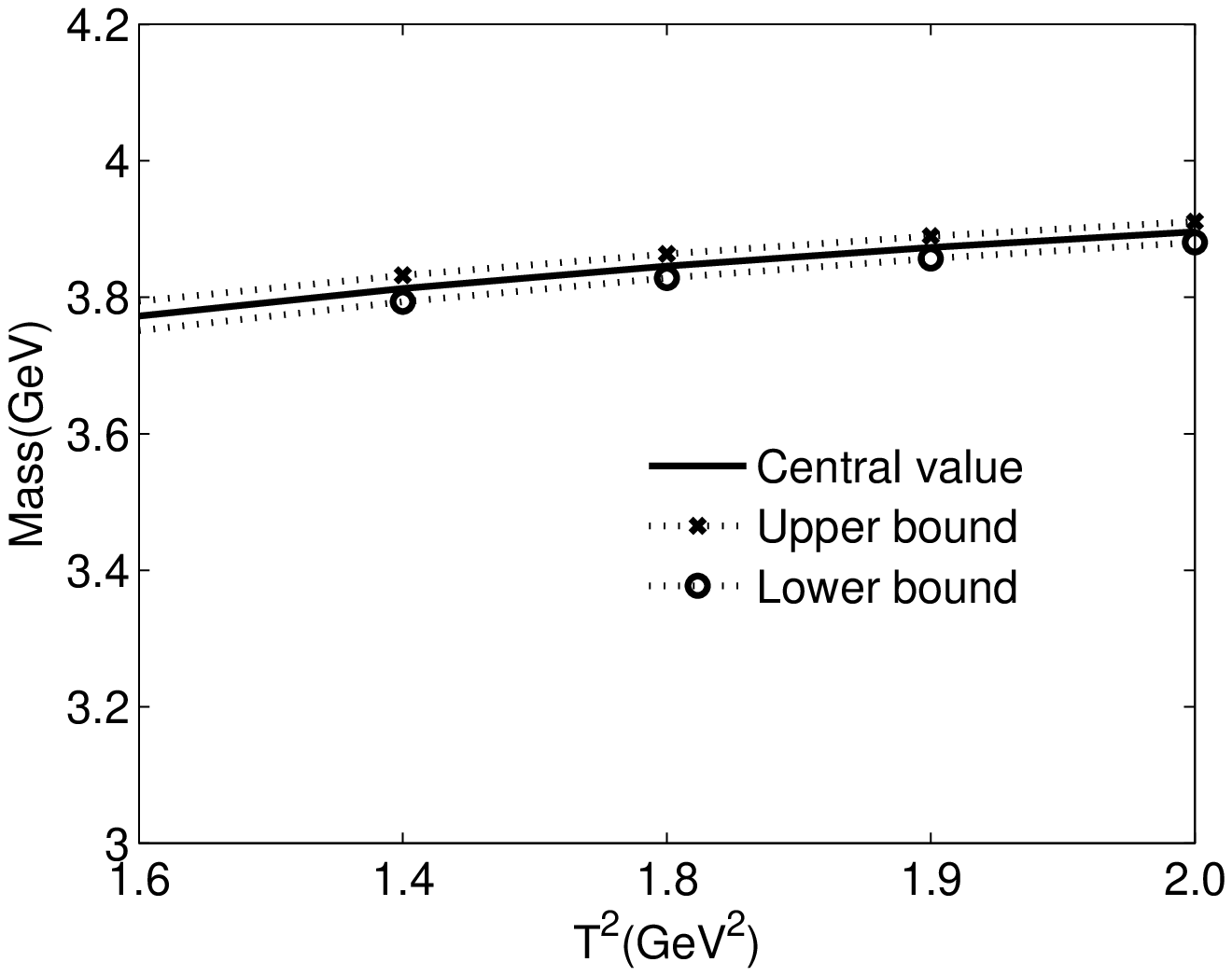}
\caption{The mass $M_{X(3842)}$ with variations of the Borel parameter $T^2$.\label{your label}}
\end{minipage}
\end{figure}
It can be seen from Fig.1 that the smaller values of the Borel parameter lead to larger pole contributions. When the values of the Borel parameter are larger
than $2.0GeV^{2}$, the pole contribution will be less than $40\%$. However, if too small values of the Borel parameters are taken(see Figs.2 and 3), the stability of
the results and the convergence of the operator product expansion can not be satisfied. If the Borel parameters are larger than $1.6GeV^2$, the
operator product expansion is well convergent. After a compromise, we choose $T^{2}=1.6\sim2.0GeV^{2}$ as our "Borel windows", which can ensure the two criteria of the QCD sum rules.
The "Borel window" is showed in Figs.4 which indicates the dependence of the mass $M_{X(3842)}$ on the Borel
parameter. The pole contribution in the "Borel window" lie in the range $40\%\sim75\%$. After taking into account the uncertainties of the input parameters, the uncertainties of
the mass $M_{X(3842)}$ are also presented in Fig.4 which are marked as the Upper bound and Lower bound. Finally, we obtain the mass of $X(3842)$,
\begin{eqnarray}
M_{X(3842)}=(3.844^{+0.0675}_{-0.0823}\pm0.020)GeV
\end{eqnarray}
where the first part of the uncertainties in the result comes from the input parameters,
and the second part originates from variations of the result in the Borel window. Considering the uncertainties of the result, the predicted mass is in excellent agreement well
with the experimental value $M_{X(3842)}=3842.71\pm0.16\pm0.12MeV$ from the LHCb collaboration. The calculations based on the QCD sum rules support assigning the $X(3842)$ to
be a $1^{3}D_{3}$ $\overline{c}c$ meson. In addition, we also calculate the decay constant of $X(3842)$ which is showed in Fig.5. We can see that the results are also stable
with variations of the Borel parameters in the "Borel window"($T^{2}=1.6\sim2.0GeV^{2}$), it is reasonable to extract the decay constant,
\begin{eqnarray}
f_{X(3842)}=(15.1374^{+0.2656}_{-0.0864}\pm0.700)GeV^{4}
\end{eqnarray}
The predicted constant $f_{X(3842)}$ can be used to study the hadronic coupling constants involving the $X(3842)$ with the three-point QCD sum rules or the light-core QCD sum rules.
\begin{figure}[h]
\begin{minipage}[t]{0.45\linewidth}
\centering
\includegraphics[height=5cm,width=7cm]{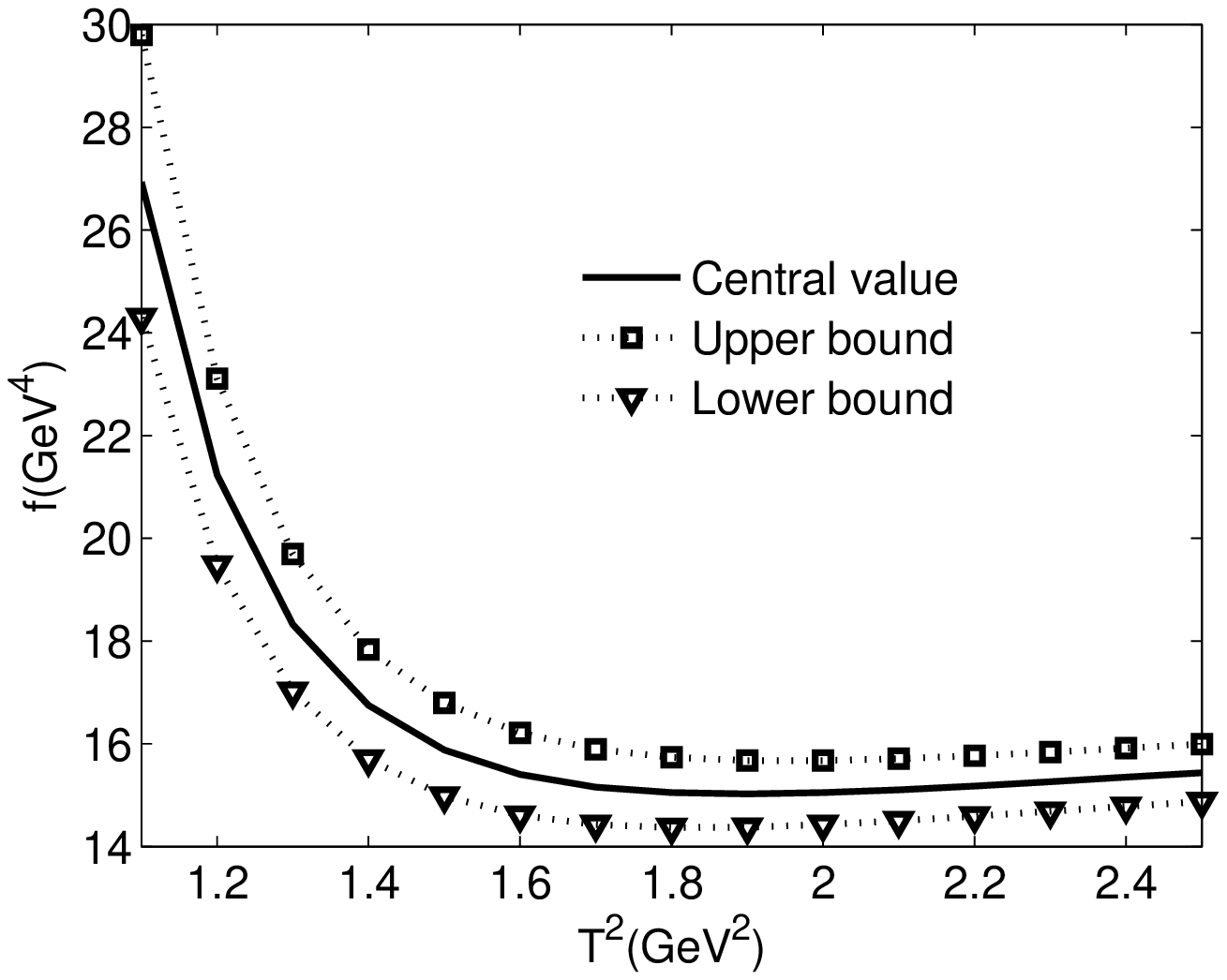}
\caption{The decay constant $f_{X(3842)}$ with variations of the Borel parameter $T^2$.\label{your label}}
\end{minipage}
\hfill
\begin{minipage}[t]{0.45\linewidth}
\centering
\includegraphics[height=5cm,width=7cm]{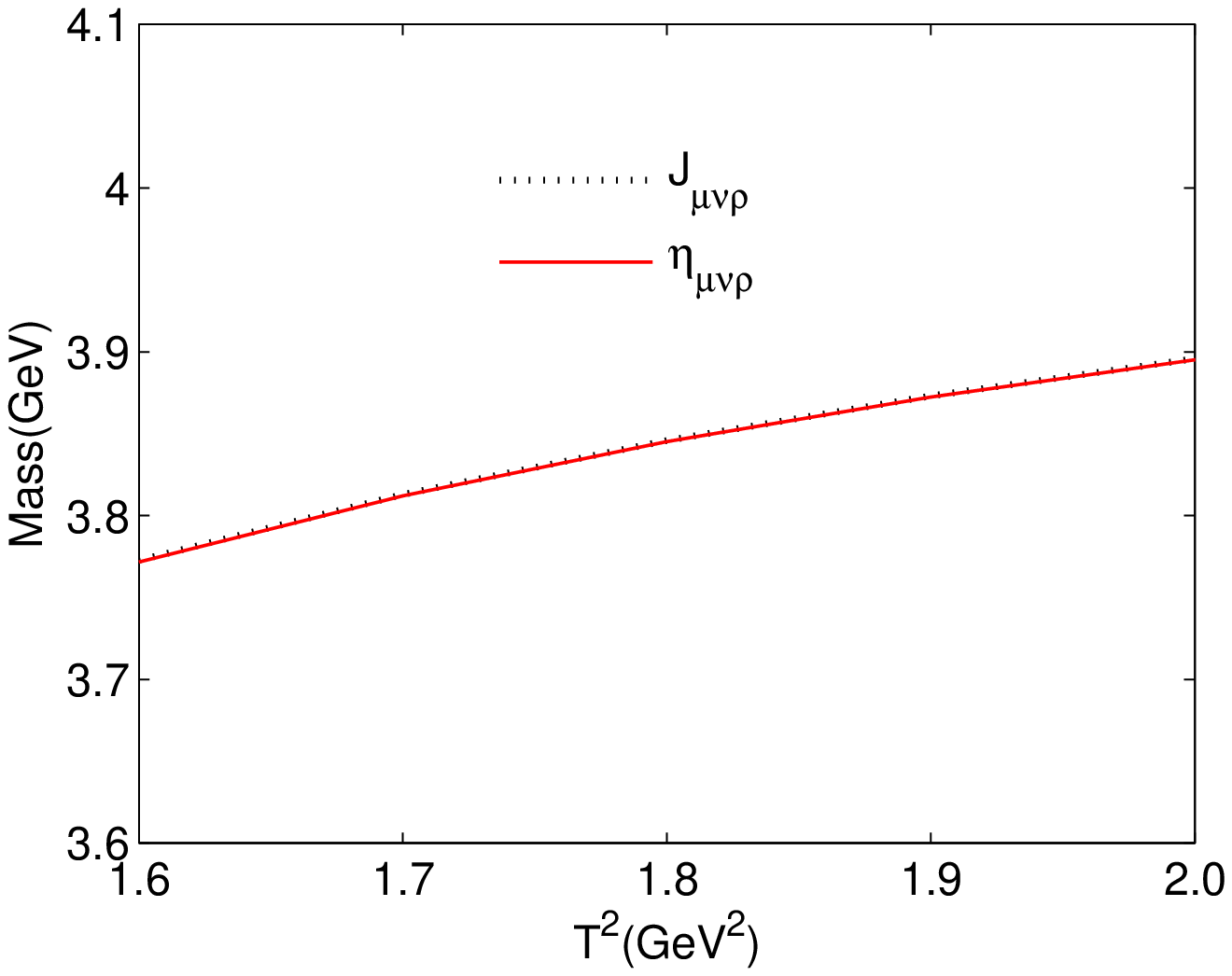}
\caption{The mass $M_{X(3842)}$ with variations of the Borel parameter $T^2$,
in which the currents $J_{\mu\nu\rho}$ and $\eta_{\mu\nu\rho}$ are considered respectively.\label{your label}}
\end{minipage}
\end{figure}

Finally, we give a simple discussion about the results which are obtained from different currents defined by Eqs(3) and (4). It can be seen from Fig.6 that different currents
$J_{\mu\nu\rho}$ and $\eta_{\mu\nu\rho}$ lead to little difference of the mass $M_{X(3842)}$ in the "Borel window". From Fig.3, we can also see that the contributions coming from
$\langle\frac{\alpha_{s}}{\pi}GG\rangle$ and $\langle g_{s}^{3}GGG\rangle$ are less than $1\%$ in the "Borel window"($T^{2}=1.6\sim2.0GeV^2$). This indicates that the contribution
of current $J_{\mu\nu\rho}^{V}$ for $X(3842)$ $\overline{c}c$ meson is too small to lead to large difference in the final results.

\begin{large}
\textbf{3 Decay properties of $X(3842)$}
\end{large}

The strong decay of $X(3842)$ will be computed using the $^{3}P_{0}$ model which was
first introduced by Micu in 1969\cite{Micu} and further developed
by other collaborations\cite{Carlitz,Yaouanc}. For now, it has
been extensively applied to evaluate the strong decays of the heavy
mesons in the charmonium\cite{Ackleh,Ferretti1,Ferretti2,Ortega} and
bottommonium systems\cite{Ferretti3,Close3,Segovia}, the
baryons\cite{ZhaoZ} and even the teraquark states\cite{LiuXW}.

In the frame work of $^{3}P_{0}$ model, a quark-antiquark pair $q\overline{q}$($q=u,d,s$) is
created from the vacuum with $0^{++}$ quantum number. With
$c$, $\overline{c}$ quarks within the initial meson $X(3842)$, this quark system
regroups into two outgoing mesons via quark rearrangement for the
meson decay process $X(3842)$$\rightarrow$$D^{+}D^{-}$ or $D^{0}\overline{D^{0}}$. Its transition operator in
the nonrelativistic limit reads
\begin{equation}
\begin{split}
    T=
    &-3\gamma\sum_{m}\langle1m1-m\mid00\rangle\int d^{3}\vec p_{3}d^{3}\vec p_{4}\delta^{3}(\vec p_{3}+\vec p_{4})\mathcal{Y}_{1}^{m}(\frac{\vec p_{3}-\vec p_{4}}{2})
    \chi_{1-m}^{34}\varphi_{0}^{34}\omega_{0}^{34}b_{3}^{\dag}(\vec p_{3})d_{4}^{\dag}(\vec p_{4})
\end{split}
\end{equation}
where $\vec{p}_{3}$ and $\vec{p}_{4}$ are the momenta of this
quark-antiquark pair$. \gamma$ is a dimensionless parameter
reflecting its creation strength. In the calculations, we commonly employ simple harmonic oscillator (SHO) approximation as the meson space
wave function
\begin{equation}
\begin{split}
\Psi_{nLM_{L}}(\vec p)=
&(-1)^{n}(-i)^{L}R^{L+\frac{3}{2}}\sqrt{\frac{2n!}{\Gamma(n+L+\frac{3}{2})}}exp(-\frac{R^{2}p^{2}}{2})L_{n}^{L+\frac{1}{2}}(R^{2}p^{2})\mathcal{Y}_{LM_{L}}(\vec
p)
\end{split}
\end{equation}
\begin{figure}[h]
\centering
\includegraphics[height=6.5cm,width=9cm]{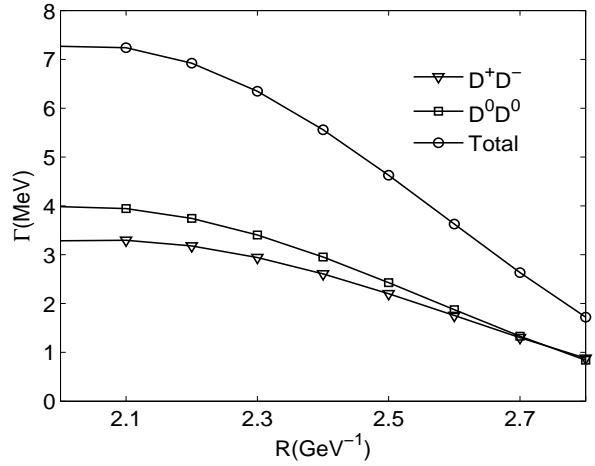}
\caption{Decay width of X(3842) as the $1^{3}D_{3}$ state with variations of the parameters R.\label{your label}}
\end{figure}
Thus, the decay width based on $^{3}P_{0}$ model depends on the following input parameters: quark pair creation strength $\gamma$ and
the SHO wave function scale parameter $R$. We take the value of $\gamma=6.3$\cite{Blunder} which is higher than that used by Kokoski and Isgur\cite{Kokoski}
by a factor of $\sqrt{96\pi}$ due to different field theory conventions, constant factors in $T$ etc. As for the scale parameter $R$, there are
mainly two kinds of choices which are the common value and the effective value. The effective
value can be fixed to reproduce the realistic root mean square radius by solving the Schrodinger
equation with a linear potential\cite{Godfrey,LiBQ5}. For the mesons $D$ and $D^{0}$, their values are taken to be $R_{D^{0}[D^{\pm}]}=1.52GeV^{-1}$\cite{Godfrey}. For a
$\overline{c}c$ system, the $R$ value of $1D$ state is estimated to be $2.3\sim2.5GeV^{-1}$\cite{YangYC}.

Taken to be a $1^{3}D_{3}$ $\overline{c}c$ meson, $X(3842)$ has only two strong decay channels $X(3842)\rightarrow D^{+}D^{-}$, $D^{0}\overline{D^{0}}$.
From Fig.7, we can clearly see the decay width of $X(3842)$ with variations of the parameter $R$.
Taking $R=2.3\sim2.5GeV^{-1}$ discussed above, the total width of $1^{3}D_{3}$ state  ranges from $5.6MeV$ to $6.9MeV$(see Fig.7).
In our previous work, we have discussed the uncertainties of the results which are predicted by $^{3}P_{0}$ decay model.
Once the optimal values of the $\gamma$ and $R$ are determined, the best predictions
based on the $^{3}P_{0}$ decay model are expected to be within a factor of $2$\cite{Blunder,gly}. More detailed analysis about
the uncertainties of the results in the $^{3}P_{0}$ decay model can be found in Ref.\cite{gly}.
Considering the uncertainties of the decay model, the calculated width is roughly compatible with the experimental data $\Gamma_{X(3842)}=(2.79\pm0.51\pm0.35)MeV$.
That is to say, it is reasonable to assign the $X(3842)$ to be the $1^{3}D_{3}$  charmonium state.
Finally, we also obtain the decay ratio $\frac{\Gamma(X(3842)\rightarrow
D^{+}D{-})}{X(3842)\rightarrow D^{0}\overline{D^{0}})}=0.85\sim0.9$ for this charmonium state. This ratio can be used to make a further confirmation about this state
in the future by LHCb collaboration.

\begin{large}
\textbf{4 Conclusion}
\end{large}

In this article, we assign the $X(3842)$ to be a D-wave $\overline{c}c$ meson, and study its mass and decay constant with the QCD sum rules. In our calculations,
we consider the contributions of the vacuum condensates up to dimension-6 in the operator product expansion. The predicted mass $M_{X(3842)}=(3.844^{+0.0675}_{-0.0823}\pm0.020)GeV$
is in agreement well with the experimental data $M_{X(3842)}=(3842.71\pm0.16\pm0.12)MeV$ from the LHCb collaboration. This result supports
assigning the $X(3842)$ to be the $1^{3}D_{3}$ $\overline{c}c$ meson. The decay constant of $X(3842)$ is predicted to be
$f_{X(3842)}=(15.1374^{+0.2656}_{-0.0864}\pm0.700)GeV^{4}$, which can be used to study the strong coupling constants involving the $X(3842)$ with the three-point QCD sum rules or the
light-cone QCD sum rules. Finally, we also calculate the strong decay width of the $1^{3}D_{3}$ state with the $^{3}P_{0}$ decay model.
Considering the uncertainties of the decay model, the calculated width is compatible with the experimental data $\Gamma_{X(3842)}=(2.79\pm0.51\pm0.35)MeV$.


\begin{large}
\textbf{Acknowledgment}
\end{large}

This work has been supported by the Fundamental Research Funds for the Central Universities, Grant Number $2016MS133$, Natural Science Foundation of HeBei Province, Grant Number $A2018502124$.

\end{document}